\documentclass[aps,reprint,superscriptaddress,nofootinbib,amsmath,amssymb]{revtex4-2}
\usepackage{mathrsfs}

\begin{document}

\title{Uniqueness of the Jackiw non-Noetherian conformal scalar field}
\thanks{We dedicate this note to Prof.\ Roman Jackiw, the pioneer
in the exploration of non-Noetherian conformal scalar fields and a
champion of subtle thinking in the wonderland of symmetries.}

\author{Eloy Ay\'on-Beato}
\email{ayon-beato-at-fis.cinvestav.mx}
\affiliation{Departamento de F\'{\i}sica, Cinvestav, Apartado Postal 14740,
07360 CDMX, M\'exico.}
\author{Mokhtar Hassaine}
\email{hassaine-at-inst-mat.utalca.cl}
\affiliation{Instituto de Matem\'atica, Universidad de Talca, Casilla 747,
Talca, Chile.}

\begin{abstract}
Jackiw was undoubtedly the first to exhibit an example of a scalar field
action which is not conformally invariant whereas its equation of motion
is. This feature has recently been dubbed as a non-Noetherian conformal
scalar field. The paradigmatic example of Jackiw was the generalization to
curved spacetime of the two-dimensional Liouville action. Here, we prove
that, up to second order, this is the unique example of a non-Noetherian
conformal scalar field in two dimensions. We establish this result using an
old and somewhat forgotten theorem which is none other than the solution to
the inverse problem of the calculus of variations.
\end{abstract}

\maketitle

One of the most wonderful equations in two dimensions is certainly
the Liouville equation \cite{Liouville}
\begin{equation}\label{eq:LiouvilleEqflat}
\Delta\phi+K e^{\phi}=0,
\end{equation}
which finds applications in various fields of physics and
mathematics. Here $\Delta$ stands for the standard two-dimensional
Laplace operator and $K$ is a constant whose interpretation can vary
according to the domain in which the Liouville equation applies. For
example, in Riemannian geometry the Liouville equation is satisfied
by the conformal factor $e^{\phi}$ of a surface of constant Gaussian
curvature $K$. One could also mention the relations existing between
the theory of Liouville and the models of Wess-Zumino-Witten as well
as those of Toda (see the excellent book \cite{FMS}). In this short
note, we will rather be interested in the symmetries underlying the
Liouville equation in curved space. Before attacking this problem,
we would like to make a few observations which will be relevant for
our purpose. First of all, the Liouville equation
(\ref{eq:LiouvilleEqflat}) in flat spacetime can be obtained from
the variation of the following two-dimensional action
\begin{equation}\label{eq:LiouvilleActionflat}
S[\phi]=\int d^2x \left(-\frac12\partial_\mu\phi\,\partial^\mu\phi+K
e^{\phi}\right).
\end{equation}
Secondly, this action and the Liouville equation \eqref{eq:LiouvilleEqflat}
are both invariant under conformal diffeomorphisms generated by a flat space
conformal Killing vector $f^{\mu}$, whose action on the scalar field is given
by
\begin{equation}\label{eq:confKilling}
\delta\phi=f^{\mu}\partial_{\mu}\phi+\partial_{\mu}f^{\mu}.
\end{equation}
Although the Liouville action \eqref{eq:LiouvilleActionflat} is
conformally invariant, its canonical energy-momentum tensor is not
traceless. Nevertheless, as shown by Jackiw \cite{Jackiw:2005su}
this tensor can be improved, by adding an extra piece ensuring the
traceless condition, as {\small
\begin{equation*}
\theta_{\mu\nu}=\partial_\mu\phi\partial_\nu\phi-\eta_{\mu\nu}\!\left(\frac12
\partial_\alpha\phi\,\partial^\alpha\phi-Ke^{\phi}\right)
+2\left(\eta_{\mu\nu}\Delta-\partial_\mu\partial_\nu\right)\phi.
\end{equation*}}%
This improved energy-momentum tensor can be shown to arise in the
flat limit of the metric variation of the Liouville action in a
curved space
\begin{equation}\label{eq:LiouvilleActioncurved}
S[\phi,g]=\int d^2x\,\sqrt{-g}
\left(-\frac12\partial_\mu\phi\,\partial^\mu\phi+K
e^{\phi}-R\phi\right),
\end{equation}
where $R$ is the scalar curvature. On the other hand, its variation
with respect to $\phi$ generalizes the Liouville equation to
\begin{equation}\label{eq:Liouvillesfeq}
\Box\phi+Ke^{\phi}-R=0,
\end{equation}
where $\Box=\nabla_{\mu}\nabla^{\mu}$ is the d'Alembertian operator in curved
spacetime. It is worth insisting that only $\theta_{\mu\nu}$ is traceless and
not the energy-momentum tensor associated to the curved Liouville action
\eqref{eq:LiouvilleActioncurved}. As a direct consequence, the curved
Liouville action is not invariant under the Weyl conformal transformations
\begin{equation}\label{eq:Weyltransf}
g_{\mu\nu} \mapsto e^{2\sigma(x)}g_{\mu\nu},\qquad \phi \mapsto
\phi-2\sigma(x),
\end{equation}
while the scalar field equation \eqref{eq:Liouvillesfeq} is, as remarkably
emphasized by Jackiw \cite{Jackiw:2005su}. This dichotomy between the
conformal invariance of the scalar field equation and not of the action has
been recently dubbed \emph{non-Noetherian conformal symmetry}
\cite{Ayon-Beato:2023bzp}. In this last reference, a full characterization of
non-Noetherian conformal scalar fields in four dimensions was provided,
generalizing the later example of Fernandes \cite{Fernandes:2021dsb},
obtained thanks to a nonminimal coupling to the Gauss-Bonnet density in
analogy with the $R\phi$ coupling present in
\eqref{eq:LiouvilleActioncurved}.

Here, we will prove that the curved Liouville action
\eqref{eq:LiouvilleActioncurved} is the unique example in two
dimensions giving rise to a scalar field equation of order two which
is invariant under the action of the conformal Weyl transformations
as given by \eqref{eq:Weyltransf}. As already emphasized in
\cite{Ayon-Beato:2023bzp}, a conformally invariant equation can
always be written by means of scalar quantities built out of an
auxiliary metric defined here by
\begin{equation}\label{eq:auxmetric}
\tilde{g}_{\mu\nu}=e^{\phi}\,{g}_{\mu\nu}.
\end{equation}
It has the advantage of being manifestly invariant under the
transformations (\ref{eq:Weyltransf}), whose infinitesimal versions
give
\begin{equation}\label{eq:infWeyltransf}
\delta_{\sigma}g_{\mu\nu}={2\sigma}g_{\mu\nu},\qquad
\delta_{\sigma}\phi=-2\sigma,\qquad\delta_{\sigma}\tilde{g}_{\mu\nu}=0.
\end{equation}
Indeed, the scalar field equation \eqref{eq:Liouvillesfeq} can be
conveniently re-written in the auxiliary frame as
\begin{equation}\label{eq:Liouvillesfeqframetilde}
-\tilde{R}+K=0,
\end{equation}
pointing out that the constant $K$ fixes now the value of the auxiliary
scalar curvature. It is also obvious that any scalar term constructed from
this auxiliary metric can be added to Eq.~\eqref{eq:Liouvillesfeqframetilde}
without breaking conformal invariance. However, the real challenge is to be
able to select only those terms that come from an action principle. In our
aim to demonstrate that action \eqref{eq:LiouvilleActioncurved} is the most
general action describing a non-Noetherian second-order conformal scalar
field, we will essentially follow the methods introduced in
\cite{Ayon-Beato:2023bzp}  by adapting them to the two-dimensional case. The
first step is to work in the auxiliary frame \eqref{eq:auxmetric}. The
generic form of a conformally invariant scalar field equation must be a
function of $\phi$ and its derivatives up to second order, together with the
auxiliary metric $\tilde{g}_{\mu\nu}$ and its associated curvatures which, in
two dimensions, reduce only to the scalar curvature. Hence, the generic
conformal equation must have the following form
\begin{equation}\label{eq:2do}
E(\phi,\tilde{\nabla}_\mu\phi,\tilde{\nabla}_{\mu}\tilde{\nabla}_{\nu}\phi,
\tilde{g}_{\mu\nu},\tilde{R})=0.
\end{equation}
The conformal invariance \eqref{eq:infWeyltransf} of this generic
equation will require that
\begin{equation}\label{eq:infinCI}
-\frac12\delta_\sigma E = \frac{\partial E}{\partial\phi}\sigma
+\frac{\partial
E}{\partial\tilde{\nabla}_\mu\phi}\tilde{\nabla}_\mu\sigma
+\frac{\partial
E}{\partial\tilde{\nabla}_\mu\tilde{\nabla}_{\nu}\phi}
\tilde{\nabla}_\mu\tilde{\nabla}_{\nu}\sigma=0,
\end{equation}
and this will be possible for any conformal factor $\sigma$ only if
the scalar quantity $E$ is independent of the scalar field and its
covariant derivatives with respect to the auxiliary metric.
Equivalently, the associated pseudoscalar which is the natural
quantity that could be derived from a covariant action must be of
the form
\begin{equation}\label{eq:2doCIps}
\mathscr{E}=\sqrt{-\tilde{g}}\,E(\tilde{g}_{\mu\nu},\tilde{R})=0.
\end{equation}
This justifies why conformally invariant equations under the
transformations \eqref{eq:infWeyltransf} can be represented in terms
of scalar quantities built out of the auxiliary metric. We must
select from all these writings \eqref{eq:2doCIps} only those which
come from a principle of action. This task can be done by using the
so-called \emph{inverse problem of the calculus of variations}
\cite{Olver:1986}. As its name suggests, the inverse problem
consists in determining if a set of equations are the Euler-Lagrange
equations of a certain action. Without entering into historical
considerations, the solution to the inverse problem of the calculus
of variations obtained by Helmholtz \cite{Helmholtz:1887} and
Volterra \cite{Volterra:1913} can be summarized as follows. The
pseudoscalar equation $\mathscr{E}=0$ (\ref{eq:2doCIps}) comes from
an action principe if and only if its Fr\'echet derivative, defined
by the operator
\begin{equation}\label{eq:FrechetE}
\delta_\phi\mathscr{E}=\sqrt{-\tilde{g}}\left(\!E-\tilde{g}^{\mu\nu}
\frac{\partial E}{\partial\tilde{g}^{\mu\nu}}
-\tilde{J}(\tilde{R}+\tilde{\Box})\right)
\!\delta\phi\equiv\mathrm{D}_\mathscr{E}(\delta\phi),
\end{equation}
is self-adjoint (see \cite{Ayon-Beato:2023bzp} for detailed
definitions). Here we abbreviate $\tilde{J}\equiv{\partial
E}/{\partial\tilde{R}}$, and a calculation of the adjoint of the
Fr\'echet derivative gives
\begin{equation}\label{eq:AdFrechetE}
\mathrm{D}_\mathscr{E}^*(A)-\mathrm{D}_\mathscr{E}(A)=
-\sqrt{-\tilde{g}}\left(A\tilde{\Box}\tilde{J} +2\tilde{\nabla}^\mu
A\tilde{\nabla}_\mu\tilde{J}\right)=0,
\end{equation}
for any function $A$. Consequently, self-adjointness is ensured provided that
$\tilde{\nabla}_\mu\tilde{J}=0$, i.e.\ the Helmholtz conditions in
two-dimensions reduce to solving the following simple equation
\begin{equation}\label{eq:Helmholtz2d}
\frac{\partial E}{\partial\tilde{R}}=\text{const.}
\end{equation}
Its general solution is an affine dependence in the auxiliary scalar
curvature that modulo a global scaling can be  written as
\eqref{eq:Liouvillesfeqframetilde}. Finally, returning to the
original frame we conclude that the only second-order field equation
conformally invariant under the transformations
\eqref{eq:Weyltransf} and coming from a principle of action in two
dimensions is the Liouville equation defined in curved space
\eqref{eq:Liouvillesfeq}.

We have conclusively established this uniqueness result by means of
the solution to the inverse problem of the calculus of variations,
with the only assumption of a non-trivial conformal weight for the
scalar field \eqref{eq:Weyltransf}. Since the corresponding action
is not conformally invariant, this implies that the curved Liouville
theory highlighted by Jackiw describes the only example of a
non-Noetherian conformal scalar field in two dimensions. This result
contrasts with the infinite family of theories allowing this
property in four dimensions \cite{Ayon-Beato:2023bzp}. For
establishing our result, we have emphasized on the non-trivial
conformal weight of the scalar field \eqref{eq:Weyltransf}, because
it is known that in two dimensions the action described by the
standard kinetic term of the scalar field is conformally invariant
provided a zero conformal weight for the scalar field, i.e.\
$g_{\mu\nu}\mapsto e^{2\sigma}g_{\mu\nu}$ and $\phi\mapsto\phi$. It
is clear that this theory escapes our analysis which is essentially
based on the auxiliary metric \eqref{eq:auxmetric}, whose conformal
invariance is due to the fact that the scalar field transforms non
trivially under conformal transformations of the metric. However,
there is little physical motivation to consider this case since it
has long been known that its effective quantum action is either
diffeomorphism invariant or conformally invariant, but never both at
the same time \cite{Polyakov}. Besides, the existence of this case
does not invalidate our results since the standard Klein-Gordon
action strictly corresponds to a Noetherian conformal symmetry.
Nevertheless, it would therefore be interesting to extend our study
to the case of conformally invariant scalar equations for which the
scalar field remains unchanged under a conformal transformation of
the metric.

\begin{acknowledgments}
This work has been partially funded by Conahcyt grant A1-S-11548 and
FONDECYT grant 1210889.
\end{acknowledgments}

\end{document}